\documentclass[12pt]{article}
\usepackage{amssymb}
\usepackage{graphics}
\def\rlx{\relax\leavevmode}
\def\inbar{\vrule height1.5ex width.4pt depth0pt}
\def\IZ{\rlx\hbox{\small \sf Z\kern-.4em Z}}
\def\IR{\rlx\hbox{\rm I\kern-.18em R}}
\def\ID{\rlx\hbox{\rm I\kern-.18em D}}
\def\IC{\rlx\hbox{\,$\inbar\kern-.3em{\rm C}$}}
\def\IN{\rlx\hbox{\rm I\kern-.18em N}}
\def\IP{\rlx\hbox{\rm I\kern-.18em P}}
\def\one{\hbox{{1}\kern-.25em\hbox{l}}}

\def\beq{\begin{equation}}
\def\eeq{\end{equation}}
\def\bea{\begin{eqnarray}}
\def\eea{\end{eqnarray}}
\def\ber{\begin{array}}
\def\eer{\end{array}}

\begin{document}

\begin{titlepage}

July 2001 \hfill {UTAS-PHYS-01-08}\\
\mbox{}\hfill{physics/0107047}

\vskip 1.6in
\begin{center}
{\Large {\bf Quantum field theory and phylogenetic branching }}
\end{center}

\normalsize
\vskip .4in

\begin{center}
P D Jarvis and J D Bashford, 
\par \vskip .1in \noindent
{\it School of Mathematics and Physics, University of Tasmania}\\
{\it GPO Box 252-37, Hobart Tas 7001, Australia }\\

\end{center}
\par \vskip .3in \noindent

\vspace{1cm}
A calculational framework is proposed for phylogenetics, using 
nonlocal quantum field theories in hypercubic geometry. Quadratic terms in the 
Hamiltonian give the underlying Markov dynamics, while higher degree 
terms represent branching events. The spatial dimension $L$ is the number 
of leaves of the evolutionary tree under consideration. Momentum 
conservation modulo ${\mathbb Z}_{2}^{\times L}$ in $L \leftarrow 1$ 
scattering corresponds to tree edge labelling using binary $L$-vectors. The bilocal 
quadratic term allows for 
momentum-dependent rate constants -- only the tree or trees
compatible with selected nonzero edge rates contribute to the 
branching probability distribution. Applications to 
models of evolutionary branching processes are discussed.

\end{titlepage}
Evolutionary processes are frequently represented as discrete or 
continuous time stationary Markov dynamics on some relevant set of system 
characters. Divergence events correspond to the 
initiation of two or more sibling processes, which each inherit the 
character probability distribution of the progenitor and then 
continue to evolve. It is the task of phylogenetic inference to deduce 
ancestral interrelationships given observed character probability distributions.  

Although the 
individual ingredients for modelling such branching trees are quite well 
understood (see for example \cite{Rodriguez,HendyPenny}), to date there is no 
overall \textit{dynamical} picture for phylogenetics. In this note we 
point out that 
existing tools from physics -- namely, quantum field theory and 
quantum many body theory when suitably interpreted 
in a stochastic context\cite{Doi,Peliti,Rittenberg} -- can provide 
both a theoretical perspective and a calculational framework. 
Below we sketch a general outline of our proposed model; details will 
be published in a separate paper. 

Consider a theory with Hamiltonian of the general form
${\mathcal H}(t) = {\mathcal H}_{0} + {\mathcal H}_{1}(t)$, with
\begin{eqnarray}
{\mathcal H}_{0} &=& \!  \sum_{x,y } \sum_{\alpha,\beta}
    {{\Psi}^{\alpha}}^{\dagger}(x)
	{{\mathcal M}_{\alpha}}^{\beta}(x\!-\!y){\Psi}_{\beta}(y),  \nonumber \\
{\mathcal H}_{1}(t) &=& \sum_{x,I}\delta(t-t_{I}) \left(
\sum_{\alpha} {W_{\alpha}^{I}}{{\Psi}^{\alpha}}^{\dagger}(x){\Psi}_{\alpha}(x)
\!-\! \sum_{\alpha, \beta, \gamma}
    {V_{\alpha \beta}^{I}}^{\gamma} 
{{\Psi}^{\alpha}}^{\dagger}(x){{\Psi}^{\beta}}^{\dagger}(x)\Psi_{\gamma}(x)
 \right), 
    	\label{eq:Hamiltonians}
\end{eqnarray}
for quantised fields $\Psi_{\alpha}(x)$ of type $\alpha = 1, \ldots, K$. 
The sum is taken over vertices of a unit hypercube $x, y \in {\mathbb 
Z}_{2}^{\times L}$, and the theory is manifestly translation 
invariant under $x \rightarrow x+a$, for $a \in {\mathbb 
Z}_{2}^{\times L}$. The interaction times $t_{I}$ are temporally 
ordered as $0=t_{0}<t_{1} < t_{2} < \ldots < t_{M}< t_{M+1}=T$ where $T$ is the total time 
for the evolution. As will be seen below, cubic interaction terms 
generate branching events, with the additional quadratic terms 
necessary to ensure that the theory is overall probability 
conserving\cite{Doi}.

Quantisation is imposed in such a way that the time evolution 
generated by the quadratic 
Hamiltonian ${\mathcal H}_{0}$ 
reproduces the standard Markov dynamics on each mode of the field.
Consider the following expansions in momentum 
space ${\mathbb Z}_{2}^{\times L}$: 
\begin{eqnarray}
	 {{\mathcal M}_{\alpha}}^{\beta}(x-y) = 
	 \lambda(x-y){M_{\alpha}}^{\beta}, & \quad &
	\lambda(z) = \sum_{k}\lambda(k) e^{i \pi k \cdot z}, \nonumber \\
\Psi_{\alpha}(x) &=& \sum_{k}
              e^{i \pi k \cdot x}c_{\alpha}(k).
			  \label{eq:Fourier}
\end{eqnarray}
Basis states of the system are Fock states of the form
\begin{equation}
	|\alpha_{1} k_{1}\; \alpha_{2} k_{2} \; \ldots \; \alpha_{N} k_{N} \rangle
=	c_{\alpha_{1}}^{\dagger}(k_{1})c_{\alpha_{2}}^{\dagger}(k_{2})
\ldots c_{\alpha_{N}}^{\dagger}(k_{N})|0\rangle,
	\label{eq:BasisStates}
\end{equation}
where the vacuum is defined as usual by the property of being 
annihilated by the modes $c_{\alpha}(k)$.
For the evolution of states $|P(t) \rangle$ under the time independent 
Hamiltonian ${\mathcal H}_{0}$, the solution of Schr\"{o}dinger's equation
\begin{equation}
\frac{d}{dt}|P(t) \rangle = -{\mathcal H}_{0}| P(t) \rangle 
\end{equation}
for evolution after time $T$, namely
\begin{equation}
| P(T) \rangle = e^{-{\mathcal H}_{0}T}| P(0) \rangle,
	\label{eq:FormalSoln0}
\end{equation}
must be computed with the help of the canonical commutation relations 
of the field. At this stage it is only necessary to impose the 
\textit{trilinear} condition\cite{Bert}
\begin{equation}
	\sum_{k}{[} c_{\alpha}^{\dagger}(k) c_{\beta}(k), 
	c_{\gamma}(\ell) {]} = {\delta^{\alpha}}_{\gamma}c_{\beta}(\ell).
	\label{eq:CanonicalCommRel}
\end{equation} 

Consider for example separable states such as
\begin{equation}
	|p(k_{1},t)\rangle \otimes |p(k_{2},t)\rangle \otimes
	\ldots \otimes |p(k_{N},t)\rangle  
	\label{eq:SeparableStates}
\end{equation}
representing a number of processes evolving in parallel, with each 
$|p(k,t)\rangle$ a single-particle state corresponding to
a probability distribution for characters of an individual process,
\begin{equation}
	|p(k,t)\rangle = \sum_{\alpha} p_{\alpha}(k,t) |\alpha k \rangle. 
	\label{eq:SinglePcleStates}
\end{equation}
With (\ref{eq:Fourier}), (\ref{eq:CanonicalCommRel}), 
\textit{either} fermionic \textit{or} 
bosonic quantisation lead to the time evolution of 
(\ref{eq:SeparableStates}) such that the probability distribution of each 
individual mode is given by the solution of the appropriate classical master equation,
\begin{equation}
	p_{\alpha}(k,T) = 
	({e^{-\lambda(k)T\cdot M})_{\alpha}}^{\beta} p_{\beta}(k,0) \equiv 
	{U(k)_{\alpha}}^{\beta}p_{\beta}(k,0).
	\label{eq:SingleModeSoln}
\end{equation}

Turning to the full, time-dependent Hamiltonian ${\mathcal H}(t) = {\mathcal 
H}_{0}+ {\mathcal H}_{1}(t)$, (\ref{eq:FormalSoln0}) must be replaced by
the time ordered exponential
\begin{equation}
	| P(T) \rangle = {\mathbb T}e^{-\int_{0}^{T}dt{\mathcal H(t)}}| P(0) 
	\rangle,
	\label{eq:FormalSoln}
\end{equation}
which in turn is expressible in the usual way as sums of multiple integrals of time-ordered
products $ \cdots{\mathcal H}(t'){\mathcal H}(t'') \cdots $.  
Consider in particular the $L \leftarrow 1$ process, and its evolution 
kernel representing the corresponding probability distribution of 
characters. Choose the distinct
outgoing momenta in some ordering to be the simple binary vectors $(0,0,\ldots,1)$, $(0,\ldots,1,0)$, 
$\ldots$, and $(1,0,\ldots,0)$ respectively.
Since momentum conservation modulo
${\mathbb Z}_{2}^{\times L}$ must hold by translation invariance,
this fixes the incoming momentum to be the maximal value
$(1,1,\ldots,1)$. The probability distribution is then a sum over 
all terms generated by the expansion of the 
time ordered exponential. 
Contributions from admissible tree diagrams are enumerated by labelling edges 
with momenta $k$, with vertices for interaction times $t_{I}$ having one incoming and 
two outgoing 
momenta $k$, $k'$, $k''$. Along edges,
the probability distribution $p_{\alpha}(k,t)$ evolves via 
(\ref{eq:SingleModeSoln}) for the appropriate time intervals 
$\Delta_{JI}=(t_{J}-t_{I})$ for $I<J$, so that the effective rate constant is
$\kappa(k) \equiv \lambda(k)\Delta_{JI}$. At 
vertices, momentum conservation ensures that 
a particular character type splits with appropriate sharing of the 
probability and type between the two subsequent edges (with momenta 
such that $k=k'+k''$). A plausible description of the divergence event is
${V_{\alpha \beta}^{I}}^{\gamma} \equiv 
{\delta_{\alpha}}^{\gamma}{\delta_{\beta}}^{\gamma}$,
which means that the two sibling processes commence evolution on their 
respective edges with characters distributed identically to that of 
their progenitor. Clearly, the model admits further generalisation to
nondiagonal or even trilocal or time-smeared interaction terms. Note that the 
additional diagonal quadratic terms in ${\mathcal H}_{1}(t)$ are necessary to 
ensure that the theory is overall probability conserving\cite{Doi} but 
do not contribute to the tree diagrams under consideration. 
The question of \textit{which} tree or trees contribute to $L \leftarrow 1$ 
scattering is encoded in the bilocal form of ${\mathcal H}_{0}$ 
(see (\ref{eq:Hamiltonians})). Only momenta $k$ corresponding 
to nonzero rate constants $\lambda(k)$ are allowed. 
For computation based on a given tree, 
it is thus possible to choose nonzero rate constants $\lambda(k)$
for selected momenta corresponding to the binary edge labelling 
unique to that tree's topology\cite{HendyPenny}. 

\begin{figure}[h]
		\centering
	\rotatebox{270}{\includegraphics{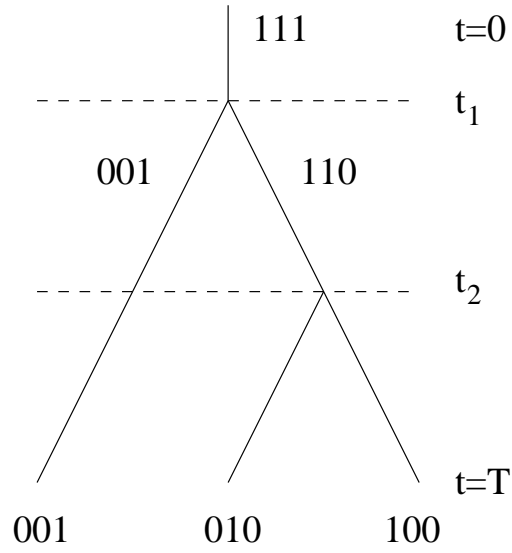}}
	\caption{Binary labelling scheme for a tree on 3 leaves ($L=3$) with 
	branching events at intermediate times $t_{1}$, $t_{2}$. Nonzero 
	rate constants for the model (\ref{eq:Hamiltonians}) are chosen for 
	the root and leaf momenta $\vec{7} = (111)$, $\vec{1} = (001)$, $\vec{2}=(010)$ and $\vec{4}=(100)$
	respectively, together with a \textit{single} additional 
momentum $\vec{6}=(110)$.}
	\label{fig:TreeFig}
\end{figure}

As an illustration, 
consider the case $L=3$, $M=2$. Nonzero 
rate constants for the model (\ref{eq:Hamiltonians}) are chosen for 
the root and leaf momenta $\vec{7} = (111)$, $\vec{1} = (001)$, $\vec{2}=(010)$ and $\vec{4}=(100)$
respectively, together with a \textit{single} additional 
momentum $\vec{6}=(110)$ (see figure 
\ref{fig:TreeFig}). Write 
${\mathcal H}_{1}(t) = {\mathcal V}^{1}\delta(t\!-\!t_{1}) + {\mathcal V}^{2}
\delta(t\!-\!t_{2})$. The time ordered 
exponential in (\ref{eq:FormalSoln}) may be written as a product, 
\begin{equation}
	{\mathbb T}e^{-\int_{0}^{T}dt{\mathcal H(t)}}=
	{\mathbb T}e^{-\int_{{2}^{+}}^{T}dt{\mathcal H(t)}}	
	V_{2}{\mathbb T}e^{-\int_{{1}^{+}}^{2^{-}}dt{\mathcal H(t)}}
	V_{1}{\mathbb T}e^{-\int_{0}^{{1}^{-}}dt{\mathcal H(t)}},
	\label{eq:TimeOrderSplit}
\end{equation}
where $V_{I}$ are time ordered exponentials for small intervals 
$\delta_{I}$ covering $t_{I}$. 
These have the form $1- {\mathcal H}_{0}\delta_{I} - {\mathcal 
V}^{I} + \cdots $, the higher order terms being ordered monomials in 
${\mathcal H}_{0}$ and ${\mathcal V}_{I}$ multiplied by nested 
$\delta$-function integrals. In the limit $\delta_{I} \rightarrow 0$,  
\begin{equation}
	{\mathbb T}e^{-\int_{0}^{T}dt{\mathcal H(t)}}=
e^{-{\mathcal H}_{0}(T\!-\!t_{2})}(1-{\mathcal V}^{2}+\cdots)e^{-{\mathcal H}_{0}(t_{2}\!-\!t_{1})}
(1-{\mathcal V}^{1}+\cdots)e^{-{\mathcal H}_{0}t_{1}}.
	\label{eq:TimeOrderWithV}
\end{equation}
Clearly the contribution to the $3 \leftarrow 1$ scattering probability 
associated with the tree of figure \ref{fig:TreeFig} is, as required, 
the unique nonzero term arising from inserting intermediate states in   
the above with the correct intermediate edge momenta, giving finally
\begin{eqnarray*}
\lefteqn{\langle \alpha_{\vec{1}} {\vec{1}}\; \alpha_{\vec{2}} {\vec{2}}\; 
	\alpha_{\vec{4}} {\vec{4}}\;|
	e^{-{\mathcal H}_{0}(T\!-\!t_{2})}{\mathcal V}^{2}e^{-{\mathcal H}_{0}(t_{2}\!-\!t_{1})}
{\mathcal V}^{1}e^{-{\mathcal H}_{0}t_{1}}|
        p(\vec{7},0) 
	\rangle = } \\
	& &\sum { U(\kappa_{\vec{2}})_{\alpha_{\vec{2}}} }^{\beta_{\vec{2}}	}
{ U(\kappa_{\vec{4}})_{\alpha_{\vec{4}}} }^{\beta_{\vec{4}}}
{V_{\beta_{\vec{2}} \beta_{\vec{4}}}^{2}}^{\gamma_{\vec{6}}}\cdot 
 { U(\kappa_{\vec{6}})_{\gamma_{\vec{6}}} }^{\beta_{\vec{6}}}
{ U(\kappa_{\vec{1}})_{\alpha_{\vec{1}}} }^{\beta_{\vec{1}}} 
{V_{\beta_{\vec{1}} \beta_{\vec{6}}}^{1}}^{\beta_{\vec{7}}} \cdot 
  { U(\kappa_{\vec{7}})_{\beta_{\vec{7}}} }^{\alpha_{\vec{7}}}  
p_{\alpha_{\vec{7}}}(\vec{7},0).
\label{ScatteringProbCalc}
\end{eqnarray*}

In phylogenetics, the probability 
distributions or dispersion tensors of characters of interest
are given directly from observations.
Whether these are compatible with calculations for a specific tree 
remains a question of statistics.
Our model (\ref{eq:Hamiltonians}) relates phylogenetic inference for 
evolutionary processes to a scattering problem for the 
associated quantum field theory.  
Recent work using Fourier-Hadamard inversion techniques for 
phylogenetic reconstruction in molecular phylogenetics\cite{Erdos,HendyPennySteel}
can be interpreted in our model as working with position states 
rather than in the momentum representation.

The overall calculational framework provided by giving 
a definite dynamical model for the branching process has potentially 
wide applicability. The picture can be extended in practice by
embellishment of various features. As already mentioned, these include
for example vertex decorations. A further possibility is a perturbative expansion of the 
quadratic term to compute the modulation of 
systematic substitution frequency types by 
the effects of Poissonian background rates. Details of the model, and 
prospects for such extensions, will be published in a separate paper.

\subsubsection*{Acknowledgements}
It is a pleasure to thank the organisers of the Second Winter 
Workshop on Mathematical Physics, University of Queensland, July 
2001, for 
providing a stimulating and multidisciplinary environment. JDB 
acknowledges the support of the University of Tasmania Institutional
Research Scheme, grant number RDOJ0011872. PDJ thanks 
David Penny and Michael Hendy for 
hospitality and discussions on a visit at the Centre for Biomolecular Sciences, 
Massey University. Finally we thank
Bob Delbourgo and Vladimir Rittenberg for fruitful comments.

\pagebreak

\end{document}